\documentclass[showpacs,aps,prd,twocolumn,floats,superscriptaddress]{revtex4}

\usepackage{graphicx}
\usepackage{amsfonts}
\usepackage{amsmath}
\usepackage{subfigure}
\usepackage{bm}
\usepackage{slashbox}

\def \ft{\widetilde}
\def \a {\alpha}
\def \Dt {\Delta t}
\def \D {\Delta}
\def \T {T_{\rm 1~day}}
\def \Om{\mathbf{\hat{\Omega}}}
\def \Dx {\mathbf{\Delta x}}
\def \eps{\varepsilon}
\def \h {{1 \over 2}}
\def\be{\begin{equation}}
\def\ee{\end{equation}}
\def\bea{\begin{eqnarray}}
\def\eea{\end{eqnarray}}
\def \no {\nonumber}
\def\lsim{\mathrel{\rlap{\lower4pt\hbox{\hskip1pt$\sim$}}
    \raise1pt\hbox{$<$}}}                
\def\gsim{\mathrel{\rlap{\lower4pt\hbox{\hskip1pt$\sim$}}
    \raise1pt\hbox{$>$}}}                

\begin{document}

\title{The cross-correlation search for a hot spot of gravitational waves}

\author{Sanjeev Dhurandhar}
\affiliation{Inter-University Centre for Astronomy and Astrophysics,\\
Post Bag 4, Ganeshkhind, Pune 411007, India}

\author{Hideyuki Tagoshi}
\affiliation{Department of Earth and Space Science,
Graduate School of Science, Osaka University, Toyonaka,
Osaka 560-0043, Japan}

\author{Yuta Okada}
\affiliation{Department of Physics, Graduate School of Science, Osaka City University,
Sugimoto 3-3-138, Sumiyoshi-ku, Osaka 558-8585, Japan}

\author{Nobuyuki Kanda}
\affiliation{Department of Physics, Graduate School of Science, Osaka City University,
Sugimoto 3-3-138, Sumiyoshi-ku, Osaka 558-8585, Japan}

\author{Hirotaka Takahashi}
\affiliation{Department of Humanities, Yamanashi Eiwa College, 888, Yokone, Kofu, Yamanashi 400-8555, Japan}
\affiliation{Earthquake Research Institute, University of Tokyo, Bunkyo-Ku, Tokyo 113-0032, Japan}
\begin{abstract}

 The cross-correlation search has been previously applied to map the gravitational wave (GW) stochastic background in the sky and also to target GW from rotating neutron stars/pulsars. Here we investigate how the cross-correlation method can be used to target a small region in the sky spanning at most a few pixels, where a pixel in the sky is determined by the diffraction limit which depends on the (i) baseline joining a pair of detectors and (ii) detector bandwidth. Here as one of the promising targets, we consider the Virgo cluster - a "hot spot" spanning few pixels  - which could contain, as estimates suggest $\sim 10^{11}$ neutron stars, of which a small fraction would continuously emit GW in the bandwidth of the detectors. For the detector baselines, we consider advanced detector pairs among LCGT, LIGO, Virgo, ET etc. Our results show that sufficient signal to noise can be accumulated with integration times of the order of a year. The results improve for the multibaseline search. This analysis could as well be applied to other likely hot spots in the sky and other possible pairs of detectors. 

\end{abstract}

\date{\today, ver.3.11}


\maketitle

\section{Introduction}
\label{intro}
An enigmatic prediction of Einstein's general theory of relativity are gravitational waves (GW). With the observed decay in the orbit of the Hulse-Taylor binary pulsar agreeing within a fraction of a percent with the theoretically computed decay from Einstein's theory, the existence of GW was firmly established. Currently there is a worldwide effort to detect GW with the operating interferometric gravitational wave observatories, the LIGO, Virgo, GEO and TAMA \cite{detectors}. Now the advanced detectors being constructed include the upgraded LIGO and Virgo, the LCGT of Japan, LIGO-Australia and future possibilities such as  Einstein Telescope (ET) \cite{fut_det}. 
\par
Different types of GW sources have been predicted and may be directly observed by these advanced detectors in the near future (see~\cite{GW_sources} and references therein for recent reviews). In this paper we will address the problem of the targeted search of stochastic GW from a small region in the sky, typically of linear size of a few degrees (few pixels - a pixel determined by the diffraction limit) - a "hot spot" - where there is likely to be an abundance of independent, unresolved GW sources continuously producing a relatively large stochastic background. Such a scenario seems feasible for the Virgo cluster, which could contain about $10^{11}$ neutron stars, the current estimate being $10^8 - 10^9$ per galaxy. Out of these neutron stars a small fraction of them could be rotating sufficiently rapidly emitting GW in the advanced detector bandwidth of several 100 Hz to about 1 kHz. These could produce a reasonable signal-to-noise ratio (SNR) with an integration time of the order of an year. Thus,  the GW source consists of spinning asymmetric neutron stars whose amplitudes and phases are randomly distributed. We will be thus dealing with a localized stochastic GW source. This is only one type of GW source, but there could be contributions from other sources such as supernovae with asymmetric core collapse, binary black hole mergers, low-mass X-ray binaries and hydrodynamical instabilities in neutron stars, or even GWs from astrophysical objects that we never knew existed. These will only in general (statistically) add to the SNR. The detectors we consider for this paper are advanced detectors such as the LIGO, Virgo, LCGT, ET etc. which are expected to have sufficient sensitivity for detecting a hot spot. 
\par
The appropriate method for observing such a source is the cross-correlation method described in \cite{mitra} (henceforth referred to as paper I), which is also generally known as the radiometric method. The idea is to cross-correlate data streams from two detectors with an appropriate time-delay, namely, the time-delay between arrival times of a GW wavefront from a specific direction $\Om$. This choice of time-delay allows the sampling of the same wavefront. As the detector baseline rotates with the earth, the time-delay between the data streams changes during the course of the day. The statistic targets a patch (pixel) in the sky around $\Om$ its size being  determined by the diffraction limit, namely, the inverse of the band-width divided by the light travel time along the baseline. This statistic in fact is a point estimate of the signal received from the given direction $\Om$ and is most appropriate for observing a hot spot and could be made optimal by `masking' the rest of the sky if the hot spot emits a strong signal. 
\par
The GW strain amplitude for a rotating neutron star is proportional to the square of the frequency \cite{JKS},
\be
h \sim 4 \pi^2 \a \frac{G}{c^4} \frac {\eps I}{R} f^2 \,,
\ee
where $\a \lsim 1$ is the orientation factor, $G$ is the Newton's gravitational constant, $c$ the speed of light, $\eps$ is the ellipticity of the neutron star, $I$ the moment of inertia, $R$ the distance to the source and $f$ the GW frequency. Since the cross-correlation  statistic is quadratic in the strain amplitude, it scales as the  fourth power of the frequency and therefore the main contribution to the SNR will tend to come from high frequency sources assuming that they are relatively abundant in the high frequency regime. Thus it is the population of millisecond neutron stars that we must primarily consider. We then estimate the millisecond neutron star population from the astrophysical information that is available and show that one can get an acceptable SNR, 
$\rho \sim 1$, for an integration of time of about an year. Using multiple baselines improves the SNR further. We find that among the current or near future baselines, the baseline of the two LIGOs and the baseline of LIGO Livingston and a LIGO like detector at AIGO site stand out - they give dominant contribution to the SNR. 
\par
In section \ref{stat}, we give a brief description of the cross-correlation method and the statistic and then derive an expression for the optimal SNR. In section \ref{R&D}, we state our results and discuss them in light of the astrophysical scenarios that are possible and the sensitivities of the future advanced detectors such as the ET.

\section{The cross-correlation statistic for targeting a hot spot}
\label{stat}

We refer to paper I for the detailed arguments involved in defining the cross-correlation statistic. Here we only furnish the salient steps. Since here we are interested in observing a hot spot, we will restrict our discussion to a point source. The full statistic, which we denote by $S$, is a weighted sum of elementary pieces $\D S_k,~ k = 1, 2, ... n$ defined over a time-segments $t_k - \Dt / 2 ~\leq~ t ~\leq~ t_k + \Dt / 2$ which are labeled by $k$. The full observation time is $T = n~\Dt$. The $\Dt$ is so chosen that it is much larger than the possible time-delay between the detectors (which must be less than about 40 ms for ground-based detectors) and much less than the time required for the orientation of the detectors to change appreciably and also on the timescale in which the noise is stationary. Current values of $\Dt$ used in LSC data analysis vary from 32 to 192 seconds.  Let us consider a pair of detectors labeled by $I = 1, 2$, then the data in the $I^{\rm th}$ detector is given by $x_I (t) = h_I (t) + n_I (t)$, the signal $h_I (t)$ is added to the noise $n_I (t)$ in the $I^{\rm th}$ detector. For a point source in the direction $\Om$, the $\D S_k$ also becomes a function of 
$\Om$. It can be expressed easily in the Fourier domain,
\begin{equation}
\D S_k (\Om)  =  \int_{-\infty}^\infty df~\ft{x}_1^*(t_k;f) \, \ft{x}_2(t_k;f) \, \ft{Q}(t_k,f, \Om) \,, 
\label{sgmt}
\end{equation}
where the $\ft{x}_I^*(t_k;f)$ are short term Fourier transforms (SFT) defined only over the interval $\Dt$ around $t_k$, namely,
\begin{equation}
 \ft{x}_I(t_k;f) \ := \ \int_{t_k-\Dt/2}^{t_k+\Dt/2} dt' \, x_I (t') \, e^{-2\pi i f t'} \,. 
\label{SFT}
\end{equation}
The $Q(t_k, f, \Om)$ is a filter function chosen so that it optimizes the filter output. It also depends on the power spectrum of the GW source and the power spectral densities of the noises in each of the detectors. As discussed in paper I, in the general case it is a far more complicated object - a functional - but for the case of a point source, it reduces to a function of the direction $\Om$. Even then it remains a functional of the signal power spectral density and the noise power spectral density (PSD). With a slight abuse of notation we still write it as a function of $f$. 
\par
The $\D S_k$ are random variables because of the noise and for different $k$ we take them to be uncorrelated. The mean and the variance of $\D S_k$ are denoted respectively by $\mu_k = \langle \Delta S_k \rangle$ and 
$\sigma_k^2 = \langle \Delta S_k^2 \rangle - \langle \Delta S_k \rangle^2$. It has been shown in paper I that the linear combination that yields the maximum SNR is:
\bea
S &=& \frac {\sum_{k=1}^n \mu_k \, \sigma_k^{-2}\, \Delta S_k}{\sum_{k=1}^n \mu_k \, \sigma_k^{-2}} \,, \\
\rho &=& \left \{ \sum_{k=1}^n \mu_k^2 / \sigma_k^2 \right \}^{\h} \,,
\label{SNR1}
\eea
where $\rho$ is the SNR. The sum over $k$ can be converted into an integral over $t$ and henceforth in this article we drop the suffix $k$ and replace $t_k$ by just $t$. This helps to avoid clutter without jeopardizing clarity.
\par
We now turn to the noise and signal PSDs in terms of which the SNR can be finally expressed. The signal cross-correlation in the two detectors in the limit of large time segment can be written as:
\begin{equation}
\langle \ft{h}^*_1(t,f) \, \ft{h}_2(t,f') \rangle \ = \ \delta(f-f') \, H(f) \, \gamma(t, f, \Om) \,, 
\label{sig}
\end{equation} 
where $\gamma(t, f, \Om)$ is the so called directed overlap reduction function analogous to the one defined in \cite{flan} for the full sky, and given in the case of the point source by,
\bea
\gamma(t, f, \Om) &=& \Gamma (t, \Om)~ e^{2\pi i f \Om\cdot\Dx(t)/c} \,,  \\ 
\Gamma (\Om, t) &=&  F_{+1}(t, \Om) F_{+2}(\Om,t) \no \\ 
&+& F_{\times 1} (\Om, t) F_{\times 2} (\Om, t) \,,
\label{Gmm}
\eea
and where the $\Dx (t)$ is the vector joining detector 1 to detector 2 and rotates with the Earth tracing out a cone. The $F_{+ I}, F_{\times I},~ I = 1, 2$, are the antenna pattern functions for the two detectors and for the two polarizations. As mentioned in paper I the directed overlap reduction function has a bandwidth of about 750 Hz as compared to the few tens of Hz for the overlap reduction function found by integrating over the full sky. This is the main advantage of this method in which the sensitive region of the detector bandwidth is sampled by the statistic. Further the quantity $f^2 H(f)$ is essentially the flux per unit frequency per unit solid angle. 
For the noise, we take the noise in the two detectors to be uncorrelated, $\langle n_1(t) n_2 (t') \rangle = 0$, and the one-sided noise PSD in each detector $I$ is given through the defining equation,
\begin{equation}
\langle \ft{n}_I^*(t;f) \, \ft{n}_I(t;f') \rangle \ = \ {1 \over 2} \, \delta(f-f') \, P_I(t;|f|). 
\label{noise}
\end{equation} 
We also assume  $\langle h_I (f) n_J (f') \rangle = 0,~I, J = 1, 2$, that is the signal and noise are uncorrelated. We are now ready to write down the optimal filter. In paper I it has been shown that the optimal filter for a given time segment labeled by $t$ and for a point source in the direction $\Om$ is given by,
\begin{equation}
Q (t, f, \Om) \ = \ \lambda ( t) \, \frac{H(f) \, \gamma^*(t, f, \Om)}{P_1(t;|f|) \, P_2(t;|f|)},
\label{mflt}
\end{equation} 
where $\lambda (t)$ is a normalization constant, which in any case cancels out in the SNR. The SNR $\rho$ is given in terms of $\mu (t)$ and $\sigma (t)$ which are the mean and standard deviation respectively of 
$\D S(t)$. To keep the expressions simple we assume that the noise in the detectors is stationary. This certainly will not be the case, but since we are only interested in order of magnitude results, the assumption is not unjustified. Then $P_I$ becomes a function of $f$ only. Also we consider a band-width $f_1 \leq f \leq f_2$ for evaluating the SNR; the lower limit $f_1$ is determined by the seismic cut-off, while the upper limit $f_2$ is decided by the GW sources above which we do not expect significant contribution to the SNR. Given this, the relevant quantities can be best expressed in terms of the following two averages:
\bea
\langle H^2 \rangle_{\rm BW} &=& \frac{2}{\D f} \int_{f_1}^{f_2} df~\frac{H^2 (f)}{P_1 (f) P_2 (f)} \,, \\
\langle \Gamma^2 \rangle_{\rm 1~day} (\Om) &=& \frac{1}{\T} \int_0^{\T} \Gamma^2 (\Om, t)~dt \,,
\label{av}
\eea 
where $\D f = f_2 - f_1$. The first is the noise weighted average of the signal $H^2 (f)$, the suffix BW denotes bandwidth, while the second is the time average of the squared directed overlap reduction function taken over one sidereal day. It is a function of sky position of the source. But since the azimuth is averaged over $2 \pi$, it is just a function of the declination of the source. Then in terms of these averages we have,
\bea
\mu (t) &=& \lambda~ (\Dt \D f)~ \langle H^2 \rangle_{\rm BW} ~ \Gamma^2 (\Om, t) \,,  \\
\sigma (t) &=& \h \lambda ~(\Dt \D f)^{\h} ~\langle H^2 \rangle_{\rm BW}^{1/2} ~\Gamma (\Om, t) \,.
\eea
Then using the continuous limit of Eq. (\ref{SNR1}), we may write the SNR $\rho$ in terms of the averages as follows:
\bea 
\rho &=& \left [\frac{1}{\Dt} \int_0^T dt~\frac{\mu^2 (t)}{\sigma^2 (t)} \right ]^{\h} \,, \no \\
&=& 2~ (T \D f)^{\h} ~ \langle H^2 \rangle_{\rm BW}^{1/2} ~ \langle \Gamma^2 \rangle_{\rm 1~day}^{1/2} \,.
\label{SNR2}
\eea
We now use this expression to compute the SNR for the continuous wave sources from the Virgo cluster. We observe that the SNR scales as $\sqrt{T}$. 
\par
To fix ideas we can look at a simplified situation of identical detectors with white noise $P_I (f) = P_0$ in the frequency range $f_1 \leq f \leq f_2$ and $P_I = \infty$ otherwise. Similarly we may consider flat signal spectrum $H (f) = H_0$, then the SNR simplifies to:
\be
\rho = 2~[T \D f]^\h ~\frac {H_0}{P_0}~ \langle \Gamma^2 \rangle_{\rm 1~day}^{1/2} \,.
\ee 

The values of $\langle \Gamma^2 \rangle_{\rm 1~day}^{1/2}$ for various combinations of detector baselines are given in Table I for the Virgo cluster which has a declination $\sim + 12.7^\circ$. 

\begin{table}[t]
\begin{center}
\begin{tabular}{|c|c|c|c|c|}
\hline
$\langle \Gamma^2 \rangle_{\rm 1~day}^{1/2}$ & LIGO-H & Virgo & LCGT & AIGO \cr \hline
LIGO-L & 0.387 & 0.288 & 0.224 & 0.452 \cr\hline
LIGO-H &  $-$  & 0.214 & 0.215 & 0.312 \cr\hline 
Virgo  &  $-$  &  $-$  & 0.276 & 0.286 \cr\hline
LCGT   &  $-$  &  $-$  &  $-$  & 0.256 \cr\hline
\end{tabular}
\caption{The values of the square root of the one day (sidereal) average of $\Gamma^2$ are given for the Virgo cluster whose declination is $\sim + 12.7^\circ$ (the RA is irrelevant since we take a one day average). LIGO-L stands for LIGO-Livingston and LIGO-H for LIGO-Hanford.}
\label{tab:gamma}
\end{center}
\end{table}

\section{Results and discussion}
\label{R&D}
\subsection{Pulsar population and distribution}
\label{sec:pulsar}
We consider gravitational waves from rotating neutron stars in the Virgo cluster. 
One important parameter of this source is the population of such neutron stars.
The number of Galactic neutron stars is estimated to be $10^8-10^9$ since the 
birth rate is about $10^{-2}/$yr and the age of the Galactic disk is about $10^{10}$yr.
What is more important in our case is the number of Galactic neutron stars 
whose rotation period is of the order of milliseconds. 
From the survey of radio pulsars in our Galactic disk,
the population of millisecond pulsars is estimated to be at least 40000 
\cite{ref:Lorimer95,ref:LorimerLRR,ref:HaiLang}
which implies a birth rate of $2.9\times 10^{-6}/$yr. 
This is consistent with other studies of the millisecond pulsar population by 
Ferrario and Wickramasinghe ($3.2\times 10^{-6}/$yr) \cite{ref:Ferrario} and 
by Story et al. ($4-5\times 10^{-6}/$yr) \cite{ref:Story}.
From the recent observation of gamma rays with the Fermi satellite \cite{ref:Abdo}, 
the population of millisecond pulsars in our Galactic globular cluster is estimated to be 
2600-4700 which is one order lower than millisecond pulsars in the Galactic disk.
Although there might be significant population of millisecond pulsars which 
do not emit radio waves, X and gamma rays now, 
since the life times of millisecond pulsars are believed to be long ($\sim 10^{10}$yr) 
\cite{ref:Camilo}, we do not expect a large population of such millisecond pulsars to exist. 
Thus we adopt $40000$ as a typical number of neutron stars per galaxy whose rotation period is of the order of milliseconds. 

\begin{figure}[t]
\begin{center}
\includegraphics[keepaspectratio,width=1\hsize]{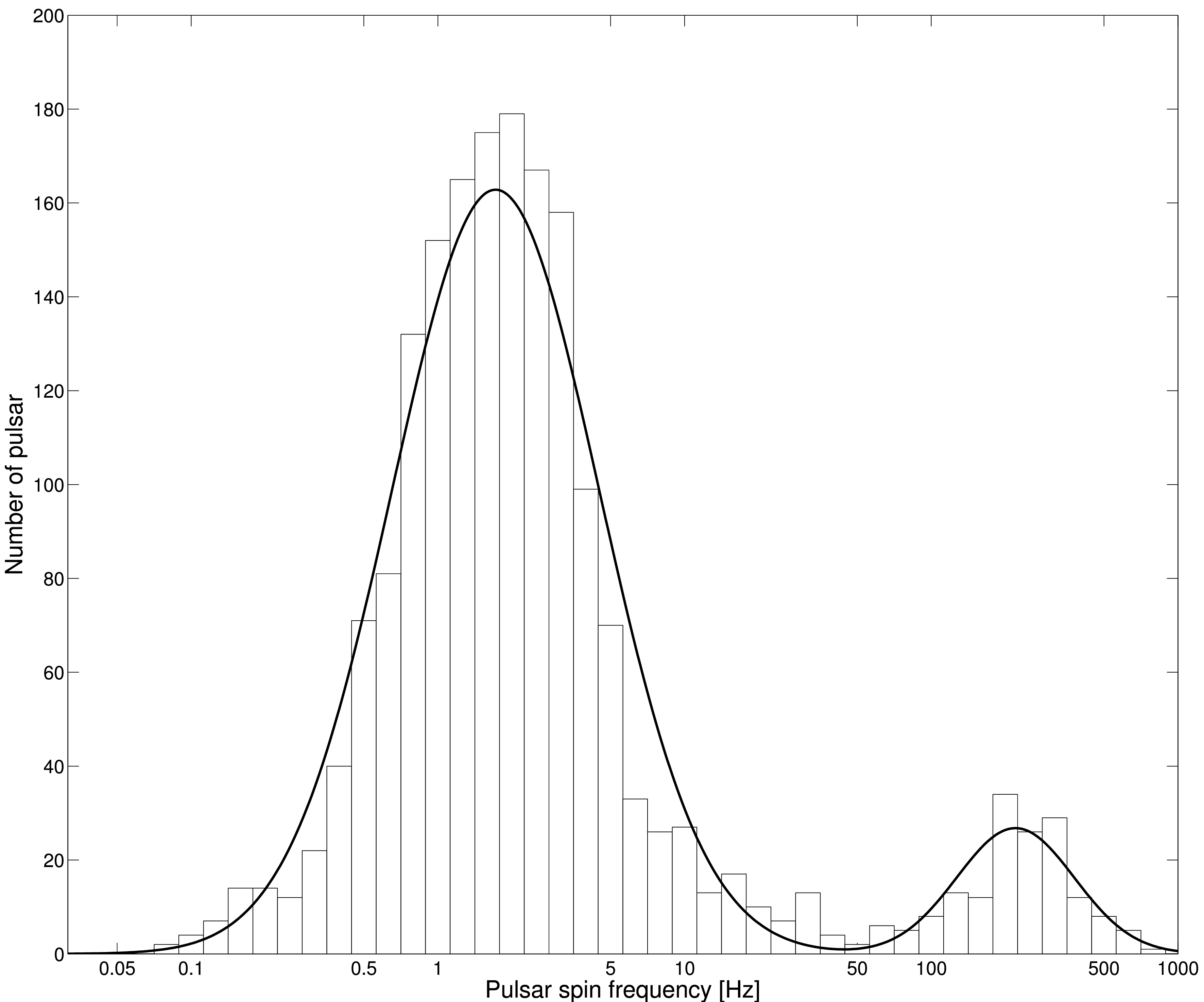}
\caption{The distribution of observed radio pulsars. 
The horizontal axis is $\log_{10}(f_r)$ where $f_r$ is the rotational 
frequency of pulsars.
The histogram is the observed number.
The solid line is the two component Gaussian model of the distribution.}
\label{fig:pulsar}
\end{center}
\end{figure}

A catalog of radio pulsars is given in the ATNF pulsar database \cite{ref:ATNF}. 
The distribution of observed radio pulsars is given in Fig.\ref{fig:pulsar}.
We find that the distribution naturally falls into two regions separated by 50Hz.
In each region, the distribution is approximately Gaussian as seen from 
the figure. This means that the distributions in each region may be approximated as 
log-normal distributions given by:
\bea
P_1(\log f_{r})d(\log f_{r})&=&\frac{1}{\sqrt{2\pi}\sigma_1}
e^{-\frac{(\log f_{r}-\log\mu_1)^2}{2\sigma_1^2}}d(\log f_{r}), \nonumber\\
&&\quad \mbox{(for $f_{r}>50$Hz)} \,, \\
P_2(\log f_{r})d(\log f_{r})&=&\frac{1}{\sqrt{2\pi}\sigma_2}
e^{-\frac{(\log f_{r}-\log\mu_2)^2}{2\sigma_2^2}}d(\log f_{r}), \nonumber\\
&&\quad \mbox{(for $f_{r}<50$Hz)} \,,
\eea
where $\mu_1=219$Hz, $\sigma_1=0.238$, $\mu_2=1.71$Hz and $\sigma_2=0.420$,
and $f_{r}=f/2$ ($f$ is the gravitational wave frequency).
$P_1$ and $P_2$ are normalized to unity when integrated from $f_{r}=0$ to infinity.
We assume a similar bimodal form of distribution of neutron stars in the Virgo cluster. 
We assume that the total number of neutron stars in our Galaxy is $10^8$ 
for $f_r<50$Hz, and $40000$ for $f_r>50$Hz. 
Since there are approximately $10^3$ galaxies in the Virgo cluster,
total number of neutron stars in Virgo cluster is $N_{\rm low}\sim 10^{11}$ 
for $f_r<50$Hz, $N_{\rm high}\sim 4\times 10^7$ for $f_r>50$Hz. 
The distribution of neutron stars including millisecond pulsars in Virgo cluster thus becomes
\bea
&&N(f)df=\nonumber\\
&& (N_{\rm high}P_1(\log f_{r})+N_{\rm low} P_2(\log f_{r}))\frac{df_r}{f_r\ln 10}.
\nonumber\\
\eea

Since the length of data of one time-segment, $\Delta t$ is at most $10^3$ seconds,
the frequency resolution is larger than $10^{-3}$ Hz. The frequency bandwidth can be taken 
as $10^3$ Hz. Thus the number of frequency bins is $10^6$. 
Since the number of pulsars with $f > 100$ Hz is $10^7$, the number of pulsars in each
frequency bin is about $10$. In the low frequency regime this number is much larger.
Thus, it is not possible to resolve the signal from each pulsar, which confirms the
stochastic nature of the Virgo cluster hot spot. 

\subsection{Signal-to-noise ratio}
\label{sec:SNR}
The spectral density of gravitational radiation from neutron stars in Virgo cluster,
$H(f)$ is given as
\bea
&&H(f)\nonumber\\
&&=\langle h^{2}\rangle N(f) \nonumber\\
&&=\left[7.05\times 10^{-34}
\left(\frac{\varepsilon}{10^{-5}}\right)
\left(\frac{I}{1.1\times 10^{45}{\rm g cm^2}}\right)
\right]^2 \nonumber\\
&&\quad 
\times\langle\alpha^2\rangle f^4 N(f) \,,
\eea
where $\langle\alpha^2\rangle$ represents the average with respect to the inclination angle
and the polarization angle. Assuming uniform distribution of the sources over the angles, 
we have $\langle\alpha^2\rangle=0.4$. We also used the distance $R=16.5$Mpc.

In order to obtain a rough idea of how large $H(f)$ is compared with the noise power 
spectral density, it is convenient to define an effective source power, $H_{\rm eff}(f)$ by,
\bea
H^2_{\rm eff}(f)=8 T_{\rm obs} \langle\Gamma^2\rangle_{\rm 1~day} f H^2(f),
\eea
where $T_{\rm obs}$ is the observation time.
\par
\begin{figure}[th]
\begin{center}
\includegraphics[width=0.9\hsize]{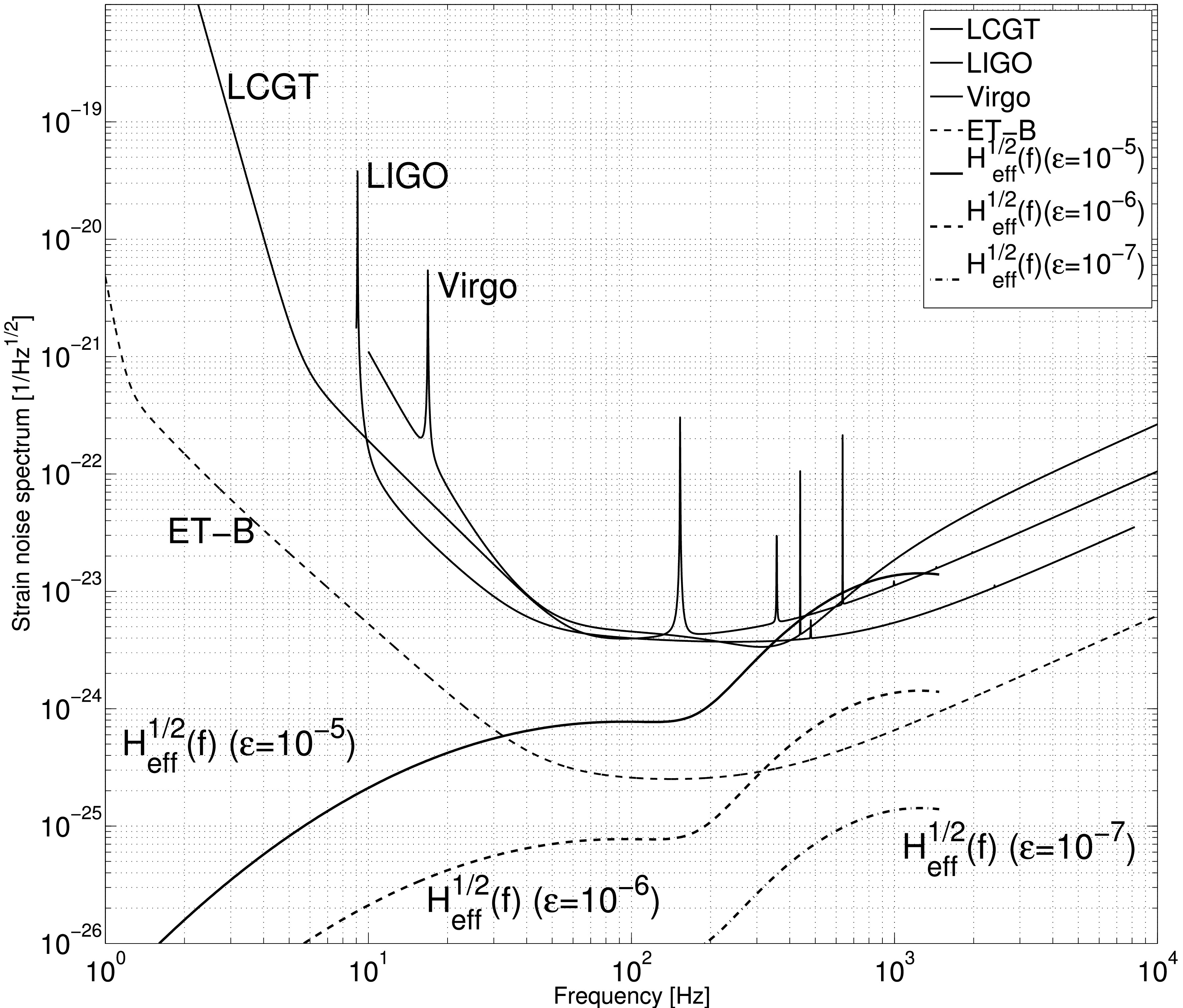}
\caption{One-sided noise power spectral density of LCGT, advanced LIGO, advanced Virgo, 
and Einstein Telescope (ET-B).
LCGT noise curve is "variable RSE in broadband mode" (VRSE(B)) \cite{ref:vBRSE}.
Advanced LIGO noise curve is "Zero Det, High Power" taken from \cite{ref:LIGOdoc}.
Advanced Virgo noise curve is take from the Virgo website\cite{ref:aVirgo}.
The Einstein Telescope noise curve is called 
"ET-B" \cite{ref:ETB}.
The effective source power $H_{\rm eff}^{1/2}(f)$ is also plotted. 
In this plot, we assume $T_{\rm obs}=1$yr and $\langle\Gamma^2\rangle_{\rm 1~day}^{1/2}=0.2$.}
\label{fig:noiseandHf}
\end{center}
\end{figure}

Then the signal-to-noise ratio is given by, 
\bea
\rho=\left [\int_{f_1}^{f_2} \frac{df}{f}~\frac{H^2_{\rm eff}(f)}{P_1 (f) P_2 (f)} \right]^{1/2}.
\eea
The noise power spectral density of various advanced detectors including Einstein Telescope,
as well as $H_{\rm eff}(f)$ are plotted in Fig.\ref{fig:noiseandHf}.
Here, we assume $T_{\rm obs}=1$yr and $\langle\Gamma^2\rangle_{\rm 1~day}^{1/2}=0.2$.
In this plot, $H_{\rm eff}(f)$ is plotted for $\varepsilon=10^{-5}, 10^{-6}, 10^{-7}$. 
Although in these plots, we include the contribution from low frequency neutron stars with $f_r<50$Hz, the contribution of these to the SNR is very small, only about a few percent.

We now consider the quantity $\langle H^2\rangle_{\rm BW}^{1/2}\Delta f^{1/2}$. We find that, 
\bea
&&\langle H^2\rangle_{\rm BW}^{1/2}\Delta f^{1/2}
\nonumber\\
&&\propto \left(\frac{\varepsilon}{10^{-5}}\right)^2
\left(\frac{I}{1.1\times 10^{45}{\rm g cm^2}}\right)^2
\left(\frac{N_{\rm msp}}{4\times 10^7}\right) \,, \nonumber\\
\eea
where $N_{\rm msp}$ is a numbers of millisecond pulsars.
The signal-to-noise ratio with 1 year observation is written as
\bea
\rho&=&\rho_{\rm 1yr} \sqrt{\frac{T_{\rm obs}}{\rm 1yr}}
\left(\frac{\langle \Gamma^2 \rangle_{\rm 1~day}^{1/2}}{0.2}\right)
\left(\frac{\varepsilon}{10^{-5}}\right)^2 \nonumber\\
&& \quad \times
\left(\frac{I}{1.1\times 10^{45}{\rm g cm^2}}\right)^2
\left(\frac{N_{\rm msp}}{4\times 10^7}\right).
\label{SNR} 
\eea
We now compute the observation time required to achieve $\rho=3$, which we denote by $T_{\rm obs}^{\rho=3}$. We choose this value of SNR because the noise in the statistic $S(\Om)$, as argued in paper I, is distributed as a Gaussian with mean $\mu_S$ and standard deviation $\sigma_S$. This is the consequence of the generalized central limit theorem. The $\rho$ is  $\mu_S / \sigma_S$. When no signal is present, we have, $\mu_S = 0$. Thus when we take $\rho >3$, there is more than 99.7\% chance that the noise is not masquerading as the signal. We then have the following result: 
\bea
T_{\rm obs}&=&T_{\rm obs}^{\rho=3}
\left(\frac{\langle \Gamma^2 \rangle_{\rm 1~day}^{1/2}}{0.2}\right)^{-2}
\left(\frac{\varepsilon}{10^{-5}}\right)^{-4} \nonumber\\
&& \quad \times
\left(\frac{I}{1.1\times 10^{45}{\rm g cm^2}}\right)^{-4}
\left(\frac{N_{\rm msp}}{4\times 10^7}\right)^{-2} \left(\frac{\rho}{3}\right)^2 \,.
\nonumber\\
\label{timeobs}
\eea

The values of $\rho_{\rm 1yr}$ and $T_{\rm obs}^{\rho=3}$ are given in 
Tables II and III for $\langle \Gamma^2 \rangle_{\rm 1~day}^{1/2} = 0.2$. These tables along with 
Eq.(\ref{SNR}) and Eq.(\ref{timeobs}) can be used to obtain the $\rho$ and the $T_{\rm obs}$ for any other value of $\langle \Gamma^2 \rangle_{\rm 1~day}^{1/2}$ or equivalently for any other sky location, not just the Virgo cluster. From the tables, we find that in the case of $\varepsilon=10^{-5}$
we can achieve $\rho=3$ in about 3 months with advanced LIGO noise PSD. 
For advanced Virgo and LCGT, it takes about 1.5 year to achieve it.
For Einstein Telescope, it will be quite easy to observe it.
However, the results strongly depend on the value of $\epsilon$. 
If $\varepsilon=10^{-6}$, it will become difficult to observe the 
Virgo cluster hot spot with advanced LIGO, advanced Virgo and LCGT.
Only Einstein Telescope will be able to detect it. Note that these are order of magnitude results where we have assumed a typical value of $\langle \Gamma^2 \rangle_{\rm 1~day}^{1/2} \sim 0.2$.

In Table IV and V, the values of $\rho_{\rm 1yr}$ and $T_{\rm obs}^{\rho=3}$ 
obtained for various detector combinations of advanced detectors, are given 
for the specific source location of the Virgo cluster.
Since $\langle\Gamma^2\rangle^{1/2}_{\rm 1~day}$ is roughly factor of 2 larger than 0.2 
for the LIGOs and AIGO network, 
the values of $\rho_{\rm 1yr}$ and $T_{\rm obs}^{\rho=3}$
are improved significantly for these baselines.
For example, the $\rho_{\rm 1yr}=3$ is achieved in 26 days by LIGO-L and LIGO-H, and 
in 19 days by LIGO-L and AIGO.

\begin{table}[thbp]
\begin{center}
\begin{tabular}{|c|c|c|c|c|}
\hline
$\rho_{\rm 1yr}$&LIGO & Virgo & LCGT & ET-B \cr \hline
LIGO  &  5.8     &  2.5  &  3.0  &  52  \cr\hline
Virgo &  $-$     &  1.5  &  1.4  &  24  \cr\hline
LCGT  &  $-$     &  $-$  &  1.6  &  27  \cr\hline
ET-B  &  $-$     &  $-$  &  $-$  &  $4.7\times 10^2$ \cr\hline
\end{tabular}
\caption{The signal-to-noise ratio $\rho_{\rm 1yr}$ which can be obtained with 1 year 
observation time for each combination of the 
detectors' noise PSD assuming $\langle \Gamma^2 \rangle_{\rm 1~day}^{\h} = 0.2$
and $\varepsilon=10^{-5}$. }
\label{tab:results1}
\end{center}
\end{table}

\begin{table}[thbp]
\begin{center}
\begin{tabular}{|c|c|c|c|c|}
\hline
$T_{\rm obs}^{\rho=3}$&LIGO & Virgo & LCGT & ET-B \cr \hline
LIGO  &  0.26~yr  &  1.5~yr  &  1.0~yr  &  1.2~day  \cr\hline
Virgo &  $-$     &  4.2~yr  &  4.7~yr  &  5.7~day  \cr\hline
LCGT  &  $-$     &  $-$    &  3.6~yr  &  4.5~day  \cr\hline
ET-B  &  $-$     &  $-$    &  $-$    &  $1.3\times 10^3$sec \cr\hline
\end{tabular}
\caption{Observation time $T_{\rm obs}^{\rho=3}$ required to achieve $\rho=3$
for each combination of the noise PSD and assuming $\langle \Gamma^2 \rangle_{\rm 1~day}^{\h} = 0.2$ and $\varepsilon=10^{-5}$.}
\label{tab:results2}
\end{center}
\end{table}

\begin{table}[thbp]
\begin{center}
\begin{tabular}{|c|c|c|c|c|}
\hline
$\rho_{\rm 1yr}$ & LIGO-H & Virgo & LCGT & AIGO \cr \hline
LIGO-L & 11.3  & 3.54 & 3.36 & 13.2 \cr\hline
LIGO-H &  $-$  & 2.63 & 3.21 & 9.12 \cr\hline 
Virgo  &  $-$  &  $-$ & 1.90 & 3.51 \cr\hline
LCGT   &  $-$  &  $-$ & $-$  & 3.84 \cr\hline
\end{tabular}
\caption{The signal-to-noise ratio $\rho_{\rm 1yr}$ which can be obtained with 1 year 
observation time for each combination of the 
detectors' noise PSD and $\langle \Gamma^2 \rangle_{\rm 1~day}^{\h}$
in Table \ref{tab:gamma}. $\varepsilon=10^{-5}$ is assumed. Noise PSD of AIGO is assumed to
be the same as that of LIGO noise PSD.}
\end{center}
\label{tab:results3}
\end{table}

\begin{table}[thbp]
\begin{center}
\begin{tabular}{|c|c|c|c|c|}
\hline
$T_{\rm obs}^{\rho=3}$[day] & LIGO-H & Virgo & LCGT & AIGO \cr \hline
LIGO-L & 25.8  & 262 & 291 & 18.9 \cr\hline
LIGO-H &  $-$  & 474 & 319 & 39.5 \cr\hline 
Virgo  &  $-$  & $-$ & 907 & 266 \cr\hline
LCGT   &  $-$  & $-$ & $-$ & 223 \cr\hline
\end{tabular}
\caption{Observation time $T_{\rm obs}^{\rho=3}$ required to achieve $\rho=3$
for each combination of the noise PSD and 
$\langle \Gamma^2 \rangle_{\rm 1~day}^{\h}$ in Table \ref{tab:gamma}. $\varepsilon=10^{-5}$ is assumed.
Noise PSD of AIGO is assumed to be the same as that of LIGO noise PSD.}
\end{center}
\label{tab:results4}
\end{table}

\begin{table}[thbp]
\begin{center}
\begin{tabular}{|ccc|}
\hline
Detector combination & $\rho_{\rm 1yr}$ & $T_{\rm obs}^{\rho=3}$[day] \cr\hline
L-H-V & 12.1    &  22.4   \cr
L-H-J & 12.2     &  22.1    \cr
L-H-A & 19.6    &  8.55    \cr
L-V-J & 5.24     &  120      \cr
L-V-A & 14.1    &  16.5    \cr
L-J-A & 14.1     &  16.4     \cr
H-V-J & 4.57    &  157     \cr
L-V-A & 10.1    &  32.1    \cr
L-J-A & 10.4     &  30.3     \cr
V-J-A & 5.54    &  107      \cr
L-H-V-J & 13.1  & 19.1     \cr
L-H-V-A & 20.4 & 7.90     \cr
L-H-J-A & 20.5  & 7.81     \cr
L-V-J-A & 15.1  & 14.4    \cr
H-V-J-A & 11.5 & 25.0     \cr
L-H-V-J-A & 21.4 & 7.21 \cr\hline
\end{tabular}
\caption{The signal-to-noise ratio $\rho_{\rm 1yr}$ which can be obtained with 1 year 
observation time and the observation time required to achieve $\rho=3$
by more than 2 detectors. These are derived from Eqs. (\ref{eq:rhonet}) and (\ref{eq:Tobsnet}) and Tables IV and V. L: LIGO-Livingston, H: LIGO-Hanford, V: Virgo, J: LCGT in Japan, A: a detector
with LIGO's noise PSD at the AIGO site in Australia.}
\end{center}
\label{tab:results5}
\end{table}

The results improve if we employ several baselines of a network of detectors. A full treatment of multi-baseline gravitational wave radiometry has been given in \cite{ref:TMB}. The results of this paper can be easily applied to the case of the hotspot where the source consists of a single pixel or at most a few pixels. In this case the beam matrix for a single baseline essentially consists of a single diagonal term for a single pixel or in case of few pixels, a small block diagonal matrix having dominant diagonal terms. The $\rho$ which we have defined above is then just the SNR obtained for the log likelihood statistic $\lambda$ defined in that paper. We also deduce from further results of that paper on sensitivity that approximately in our case, 
\be
\rho_{\rm network}^2 = \sum_{{\cal I}} \rho_{{\cal I}}^2 \,, \label{eq:rhonet}
\ee
where $\rho_{\rm network}$ is the SNR for the network and the index ${\cal I}$ runs over all the baselines of the network. Similarly, it is easy from the foregoing to deduce that the observation times to reach an SNR of 3, namely $T_{\rm obs}^{\rho = 3}$, add harmonically; more specifically we have,
\be
\frac{1}{T_{\rm obs}^{\rho = 3}} = \sum_{{\cal I}} \frac{1}{T_{\rm obs}^{\rho_{{\cal I}} = 3}} \,,
\label{eq:Tobsnet}
\ee
where now the $T_{\rm obs}^{\rho = 3}$ denotes the time of observation required for the network and 
$T_{\rm obs}^{\rho_{{\cal I}} = 3}$ denotes the observation time required for the baseline ${\cal I}$ to reach the SNR of 3. 
\par
We can now apply these results to various networks. 
The results are given in Table VI. 
We first consider the 3 detectors, LIGO-Virgo (LHV) network. Just comparing tables IV and VI, the $\rho_{\rm 1yr}$ goes up from 11.3 for two LIGOs to 12.1 for L-H-V network which is about 7 \% increase. Note that one must here take into account 3 baselines L-H, L-V and H-V. The $T_{\rm obs}^{\rho = 3}$ comes down from 25.8 days for the two LIGOs to 22.4 day for the L-H-V network which is a decrease of 13 \%. If one considers the two LIGOs along with the LCGT the improvement is almost similar to Virgo case, that is, the observation time comes down to 22.1  day. The improvement of adding other baselines to the L-H baseline is marginal because the LH contribution is dominant. Note however that an interesting improvement is obtained if one considers a detector at AIGO site  assuming same noise PSD as the LIGOs. In such a L-H-A network, 
L-A contribution becomes dominant because of largest $\langle \Gamma^2 \rangle_{\rm 1~day}^{\h}$
in Table I, and $\rho_{\rm 1yr}$ goes up to 19.6 and $T_{\rm obs}^{\rho = 3}$ comes down to 8.5 days. 
L-V-A and L-J-A networks are similar and gives second largest value of $\rho_{\rm 1yr}$ 
among 3 detector networks. They are better than L-H-V and L-H-J cases. 

In case of a 4 or 5 detector network, we can have further improvement, but 
the effect is not so large since the L-H-A contribution dominates $\rho$. 
For the 4 detector case, L-H-J-A network gives the largest value of $\rho_{\rm 1yr}=20.5$. 
L-H-V-A network also gives similar results. In case of the 5 detector network, $\rho_{\rm 1yr}=21.4$ and $T_{\rm obs}^{\rho = 3}=7.21$ days.

\section{Summary}

 In this article we address the question of observing a hotspot of stochastic GW using the cross-correlation statistic. The idea is to restrict the statistic to a single or few pixels in the sky and target possible point stochastic sources. A possible source which we pick is the Virgo cluster which could be a rich bed of rotating neutron stars containing an estimated number of $10^{11}$. Out of these the rotating neutron stars emitting GW which fall into the bandwidth of the advanced detectors are primarily the millisecond neutron stars. We assume that the distribution of such neutron stars follows a bimodal distribution similar to that of the radio millisecond pulsars observed in our galaxy. We then see that with advanced detectors the observation time required to accumulate SNR $\sim 3$ is about an order of an year if the average ellipticity of neutron stars is $\varepsilon\sim 10^{-5}$. Several baselines have been considered as well as multiple baselines corresponding to networks of detectors. In these calculations, the baselines that stand out are the two LIGO detectors and the LIGO Livingston and a LIGO like detector at the AIGO site in Australia. These baselines have the best sensitivity, because for these baselines, the  detectors are almost co-aligned. In such cases, the observation time required to achieve SNR $\sim 3$ is about 20 days if we assume $\varepsilon=10^{-5}$. The future proposed Einstein Telescope can easily detect the hotspot. In fact, Einstein Telescope would be the only detector which can observe the Virgo cluster hotspot if $\varepsilon\sim 10^{-6}$. In that case, only one Einstein Telescope will be sufficient to detect the Virgo cluster by cross-correlating with other detectors like LIGOs, Virgo and LCGT.   
 
Besides the Virgo cluster, there could be other candidates for hotspots such as the Andromeda galaxy or our own  galactic centre. Although in these cases, the number of sources contributing to the GW background may be smaller than the Virgo cluster, their distances are much smaller, which makes up for the overall strength of the stochastic sources.       

\begin{acknowledgments}

S. Dhurandhar thanks S. Bose and S. Mitra for useful discussions on multiple baselines.
We thank F. Takahara and S.J. Tanaka for useful discussions on the population of neutron stars. 
S. Dhurandhar also 
acknowledges the DST and JSPS Indo-Japan international cooperative programme 
for scientists and engineers for supporting visits to Osaka City University,
Japan and Osaka University, Japan. 
H. Tagoshi, N. Kanda and H. Takahashi thank JSPS and DST under the same Indo-Japan programme for their visit to IUCAA, Pune, India.
H.Tagoshi's work was also supported in part by a Monbu Kagakusho Grant-in-aid
for Scientific Research of Japan (No. 20540271).
H.Takahashi's work was also supported in part by a Monbu Kagakusho Grant-in-aid
for Scientific Research of Japan (No. 23740207). 
\end{acknowledgments}

\end{document}